\documentclass[prd,twocolumn,groupedaddress,showpacs,nofootinbib]{revtex4}
\usepackage{graphicx}
\usepackage{dcolumn}
\usepackage{amssymb}
\usepackage{mathrsfs}
\usepackage{amsmath}
\usepackage{epsfig}
\usepackage[dvips]{color}
\usepackage{hhline}

\begin{document}

\title{Parity: from strong CP problem to dark matter, neutrino masses and baryon asymmetry}

\author{Pei-Hong Gu}
\email{peihong.gu@mpi-hd.mpg.de}

\affiliation{Max-Planck-Institut f\"{u}r Kernphysik, Saupfercheckweg
1, 69117 Heidelberg, Germany}

\begin{abstract}

We show that in an $SU(3)^{}_c\times [SU(2)^{}_L\times
U(1)^{}_Y]\times [SU(2)'^{}_R\times U(1)'^{}_Y]$ framework, the
parity symmetry motivated by solving the strong CP problem without
resorting to an axion can predict a dark matter particle with a mass
around $302\,\textrm{GeV}$. This dark matter candidate can be
directly detected in the presence of a $U(1)^{}_Y\times U(1)'^{}_Y$
kinetic mixing. Furthermore, our model can accommodate a natural way
to simultaneously realize an inverse-linear seesaw for neutrino
masses and a resonant leptogenesis for baryon asymmetry.

\end{abstract}


\maketitle

\section{Introduction}

The most popular solution to the so-called strong CP problem is to
introduce a continuous Peccei-Quinn \cite{pq1977} symmetry which
predicts a pseudo Goldstone boson, the axion
\cite{pq1977,weinberg1978,kim1979,dfs1981}, and hence a dynamical
strong CP phase. However, the axion has not been experimentally seen
so far. Alternatively, we can consider certain discrete symmetries
to suppress or remove the strong CP phase. For example, Babu and
Mohapatra \cite{bm1989} proposed an $SU(3)^{}_c\times
SU(2)^{}_L\times SU(2)^{}_R\times U(1)^{}_{B-L}$ left-right
symmetric model \cite{ps1974} with parity symmetry to generate a
tiny strong CP phase at two-loop level. Barr, Chang and
Senjanovi\'{c} \cite{bcs1991} then pointed out the Babu-Mohapatra
scheme could be generalized in a relaxed $SU(3)^{}_c\times
[SU(2)^{}_L\times U(1)^{}_Y]\times [SU(2)'^{}_R\times U(1)'^{}_Y]$
framework. Such dark parity extension of the ordinary electroweak
theory could result in some stable dark particles, among which the
stable dark quark is not allowed to have a significant relic
density. To dilute the stable dark quark in the early universe, one
may expect an inflation with a reheating temperature much below the
mass of the stable dark quark \cite{bcs1991}.

In this paper we shall demonstrate a novel $SU(3)^{}_c\times
[SU(2)^{}_L\times U(1)^{}_Y]\times [SU(2)'^{}_R\times U(1)'^{}_Y]$
model with some interesting implications of the parity symmetry
motivated by solving the strong CP problem on the property of dark
matter, the origin of neutrino masses and the generation of baryon
asymmetry. In our model, after the dark electromagnetic symmetry is
spontaneously broken, a dark up quark can have a fast decay into a
dark electron and four ordinary fermions through the mediations of
three colored scalars. Our model also contains three gauge-singlet
fermions with small Majorana masses and an $[SU(2)^{}_L\times
SU(2)'^{}_R]$-bidoublet Higgs scalar with seesaw-suppressed vacuum
expectation value (VEV). The singlet fermions and the dark neutrinos
can form three pairs of quasi-degenerate Majorana fermions to
realize a resonant \cite{fps1995,pilaftsis1997} leptogenesis
\cite{fy1986,fnr2013} for baryon asymmetry. Meanwhile, the neutrino
masses can have a form of inverse \cite{mv1986} and linear
\cite{barr2003} seesaw \cite{minkowski1977,mw1980}. The dark
electron can annihilate into the dark photon and then obtain a relic
density to explain the dark matter puzzle. The dark matter relic
density will only depend on the dark electron mass since the parity
symmetry identifies the dark gauge couplings to the ordinary ones.
We hence can determine the dark matter mass to be about
$302\,\textrm{GeV}$ from the measured dark matter relic density. In
the presence of a $U(1)$ kinetic mixing, our dark matter particle
can be verified by the ongoing and future dark matter direct
detection experiments.

\section{The model}

The ordinary and dark scalars are denoted by
\begin{eqnarray}\begin{array}{lcl}
\phi(\textbf{1})(\textbf{2},-\frac{1}{2})(\textbf{1},0)&\leftrightarrow&
\phi'(\textbf{1})(\textbf{1},0)(\textbf{2},-\frac{1}{2})\,,\\
[2mm]
\eta(\textbf{1})(\textbf{1},-\frac{1}{3})(\textbf{1},0)&\leftrightarrow&
\eta'(\textbf{1})(\textbf{1},0)(\textbf{1},-\frac{1}{3})\,,\\
[2mm]
\delta(\textbf{3})(\textbf{1},-\frac{1}{3})(\textbf{1},0)&\leftrightarrow&
\delta'(\textbf{3})(\textbf{1},0)(\textbf{1},-\frac{1}{3})\,,\\
[2mm]
\omega(\textbf{8})(\textbf{1},-\frac{1}{3})(\textbf{1},0)&\leftrightarrow&
\omega'(\textbf{8})(\textbf{1},0)(\textbf{1},-\frac{1}{3})\,,
\end{array}
\end{eqnarray}
while the three generations of ordinary and dark fermions are
\begin{eqnarray}
\begin{array}{lcl}
q^{}_{Li}(\textbf{3})(\textbf{2},+\frac{1}{6})(\textbf{1},0)&\leftrightarrow&
q'^{}_{Ri}(\textbf{3})(\textbf{1},0)(\textbf{2},+\frac{1}{6})\,,\\
[2mm]
d^{}_{Ri}(\textbf{3})(\textbf{1},-\frac{1}{3})(\textbf{1},0)&\leftrightarrow&
d'^{}_{Li}(\textbf{3})(\textbf{1},0)(\textbf{1},-\frac{1}{3})\,,\\
[2mm]
u^{}_{Ri}(\textbf{3})(\textbf{1},+\frac{2}{3})(\textbf{1},0)&\leftrightarrow&
u'^{}_{Li}(\textbf{3})(\textbf{1},0)(\textbf{1},+\frac{2}{3})\,,\\
[2mm]
l^{}_{Li}(\textbf{1})(\textbf{2},-\frac{1}{2})(\textbf{1},0)&\leftrightarrow&
l'^{}_{Ri}(\textbf{1})(\textbf{1},0)(\textbf{2},-\frac{1}{2})\,,\\
[2mm]
e^{}_{Ri}(\textbf{1})(\textbf{1},-1)(\textbf{1},0)&\leftrightarrow&
e'^{}_{Li}(\textbf{1})(\textbf{1},0)(\textbf{1},-1)\,.
\end{array}
\end{eqnarray}
As for the $[SU(2)^{}_L\times SU(2)'^{}_R]$-bidoublet scalar and the
three gauge-singlet fermions, they are defined as
\begin{eqnarray}
\Sigma(\textbf{1})(\textbf{2},-\frac{1}{2})(\mathbf{\bar{2}},+\frac{1}{2})
\leftrightarrow\Sigma^\dagger_{}\,,
~\chi_{Ri}^{}(\textbf{1})(\textbf{1},0)(\textbf{1},0)\leftrightarrow\chi_{Ri}^{c}\,.&&
\nonumber\\
&&
\end{eqnarray}
In the above notations, the first, second and third parentheses
following the fields are the quantum numbers under the $SU(3)^{}_c$,
$SU(2)^{}_L\times U(1)^{}_Y$ and $SU(2)'^{}_R\times U(1)'^{}_Y$
gauge groups, respectively.

For simplicity, we will not write down the full Lagrangian. Instead,
we firstly show the following scalar interactions,
\begin{eqnarray}
\label{potential} V&\supset&\mu_{\phi^{}_{}}^2
\phi^\dagger_{}\phi^{}_{}+\mu_{\phi'^{}_{}}^2
\phi'^\dagger_{}\phi'^{}_{}+ \mu_{\eta^{}_{}}^2
\eta^\ast_{}\eta^{}_{}+\mu_{\eta'^{}_{}}^2
\eta'^\ast_{}\eta'^{}_{}+M_{\delta}^2\delta^\dagger_{}\delta\nonumber\\
&&+M_{\delta'}^2\delta'^\dagger_{}\delta'+M_{\omega}^2\omega^\dagger_{}\omega
+M_{\omega'}^2\omega'^\dagger_{}\omega'+M_{\Sigma}^2\textrm{Tr}(\Sigma^\dagger_{}\Sigma)\nonumber\\
&&+\lambda\delta^\dagger_{}\delta'^\dagger_{}(\omega\eta'+\omega'\eta)
+\kappa(\eta^\ast_{}\omega^\dagger_{}\delta^2_{}+\eta'^\ast_{}\omega'^\dagger_{}\delta'^2_{})\nonumber\\
&& +\xi \delta^\dagger_{}\delta'\eta\eta'^\ast_{} + \rho
\phi^\dagger_{}\Sigma \phi +\textrm{H.c.}\,.
\end{eqnarray}
where the parity symmetry is assumed to be softly broken, i.e.
\begin{eqnarray}
\mu_{\phi^{}_{}}^2\neq
\mu_{\phi'^{}_{}}^2\,,~~\mu_{\eta^{}_{}}^2\neq
\mu_{\eta'^{}_{}}^2\,,~~M_{\delta}^2 \neq
M_{\delta'}^2\,,~~M_{\omega}^2\neq M_{\omega'}^2\,.
\end{eqnarray}
Such soft breaking may arise from a spontaneous parity violation
\cite{cmp1984}. We then give all of the parity-invariant Yukawa
couplings,
\begin{eqnarray}
\label{yukawa} \mathcal{L}\!\!\!&\supset&\!\!\!
-y^{}_{\delta}(\delta\bar{u}^{}_R e^c_R+\delta'\bar{u}'^{}_L
e'^c_L)-y'^{}_{\delta}(\delta\bar{d}^{}_R
\chi^c_R+\delta'\bar{d}'^{}_L
\chi^{}_R)\nonumber\\
&&-y''^{}_{\delta}(\delta\bar{q}^{}_L
i\tau_2^{}l^c_L+\delta'\bar{q}'^{}_R
i\tau_2^{}l'^c_R)-y^{}_{\omega}(\omega\bar{u}^{c}_R
d^{}_R+\omega'\bar{u}'^{c}_L
d'^{}_L)\nonumber\\
&&-y'^{}_{\omega}(\omega\bar{q}^{c}_L i\tau_2^{}
q^{}_L+\omega'\bar{q}'^{c}_R i\tau_2^{}
q'^{}_R)-y^{}_d(\bar{q}^{}_L\tilde{\phi} d^{}_R +
\bar{q}'^{}_R\tilde{\phi}'
d'^{}_L)\nonumber\\
&&-y^{}_u(\bar{q}^{}_L\phi u^{}_R + \bar{q}'^{}_R\phi'
u'^{}_L)-y^{}_e(\bar{l}^{}_L\tilde{\phi}^{}_d e^{}_R +
\bar{l}'^{}_R\tilde{\phi}'^{}_d e'^{}_L)\nonumber\\
&&-h(\bar{l}^{}_L\phi^{}_{} \chi^{}_R + \bar{l}'^{}_R\phi'^{}_{}
\chi^{c}_R)-f\bar{l}^{}_L\Sigma l'^{}_R+\textrm{H.c.}\,,
\end{eqnarray}
where a baryon number conservation has been assumed to forbid the
other terms involving the leptoquark scalars $\delta$ and $\delta'$.
We also set the Majorana mass term of the singlet fermions
$\chi_R^{}$ as below,
\begin{eqnarray}
\mathcal{L}\supset -
\frac{1}{2}\mu\bar{\chi}^c_R\chi_R^{}+\textrm{H.c.}~~\textrm{with}~~\mu^T_{}=\mu\,.
\end{eqnarray}
Furthermore, we introduce a kinetic mixing between the $U(1)^{}_Y$
and $U(1)'^{}_Y$ gauge fields,
\begin{eqnarray}
\mathcal{L}\supset -
\frac{\epsilon}{2}B_{\mu\nu}^{}B'^{\mu\nu}_{}\,,
\end{eqnarray}
where $B_{\mu\nu}^{}$ and $B'^{\mu\nu}_{}$ are the $U(1)^{}_Y$ and
$U(1)'^{}_Y$ field strength tensors.

\section{Symmetry breaking}

As a result of the softly broken parity, the dark Higgs doublet
$\phi'$ can develop a VEV different from that of the ordinary Higgs
doublet $\phi$ to drive the dark and ordinary electroweak symmetry
breaking at different scales,
\begin{eqnarray}
\label{vev1} \langle\phi'^{}\rangle\neq \langle\phi^{}\rangle\simeq
174\,\textrm{GeV}\,.
\end{eqnarray}
Note the VEVs $\langle\phi'^{}\rangle$ and $\langle\phi^{}\rangle$
are both real. After the above symmetry breaking, the heavy Higgs
bidoublet $\Sigma$ can pick up a seesaw-suppressed VEV,
\begin{eqnarray}
\label{vev2}
\langle\Sigma\rangle&=&-\frac{\rho\langle\phi^{}_{}\rangle\langle\phi'^{}_{}\rangle}{M_{\Sigma}^2}\ll
\langle\phi^{}_{}\rangle\,,\langle\phi'^{}_{}\rangle\,.
\end{eqnarray}
Here we have rotated the cubic coupling $\rho$ to be real. Under the
softly broken parity, the dark Higgs singlet $\eta'$ can acquire a
nonzero VEV to spontaneously break the dark electromagnetic symmetry
although its ordinary partner $\eta$ is only allowed to have a zero
VEV.

By making a non-unitary transformation \cite{fh1991},
\begin{eqnarray}
\tilde{B}^{}_\mu=B^{}_\mu+\epsilon B'^{}_\mu\,,
~~\tilde{B}'^{}_\mu=\sqrt{1-\epsilon^2}B'^{}_\mu\,,
\end{eqnarray}
we can remove the $U(1)$ kinetic mixing and then define the
following orthogonal fields,
\begin{eqnarray}
\begin{array}{l}
A^{}_\mu =W^3_\mu s_W^{} +\tilde{B}^{}_\mu c_W^{}\,,~~\, Z^{}_\mu
=W^3_\mu c_W -\tilde{B}^{}_\mu
s^{}_W\,,\\
[2mm]A'^{}_\mu =W'^3_\mu s^{}_{W} +\tilde{B}'^{}_\mu c^{}_{W}\,,~~
Z'^{}_\mu =W'^3_\mu c^{}_{W} -\tilde{B}'^{}_\mu
s^{}_{W}\,.\end{array}
\end{eqnarray}
Here $s_W^{}=\sin\theta_W^{}$ and $c_W^{}=\cos\theta_W^{}$ with
$\theta_W^{}$ being the Weinberg angle, while $W^{3}_{}$ and
$W'^{3}_{}$ are the diagonal components of the $SU(2)^{}_L$ and
$SU(2)'^{}_L$ gauge fields. In the above orthogonal base, the field
$A$ is exactly massless and is the ordinary photon, according to the
unbroken electromagnetic symmetry in the ordinary sector, while the
dark photon $A'$ mixes with the $Z$ and $Z'$ bosons. As the dark
electromagnetic symmetry now is broken by the VEV
$\langle\eta'\rangle$, the dark photon can have a mass
\begin{eqnarray}
m_{A'}^{}&\simeq&
\frac{2\sqrt{2\pi\alpha}}{3}\langle\eta'\rangle\simeq1\,\textrm{GeV}
\left(\frac{\langle\eta'\rangle}{7\,\textrm{GeV}}\right)\nonumber\\
&&\textrm{for}~~\epsilon\ll1\,,
~~\langle\eta'\rangle\ll\langle\phi'\rangle\,,
\end{eqnarray}
with $\alpha=e^2_{}/(4\pi)\simeq 1/137$ being the fine structure
constant. The dark photon can couple to the ordinary fermions
besides the dark fermions \cite{gu2012-2},
\begin{eqnarray}
\mathcal{L}&\supset& eA'^{}_\mu\left\{\frac{\epsilon
}{4}\left[\bar{e}\gamma^\mu_{}
\left(3+\gamma_5^{}\right)e+\bar{\nu}\gamma^\mu_{}(1-\gamma_5^{})
\nu\right.\right.\nonumber\\
&&\left.+\bar{d}\gamma^\mu_{}
\left(\frac{1}{3}+\gamma_5^{}\right)d-\bar{u}\gamma^\mu_{}
\left(\frac{5}{3}+\gamma_5^{}\right)u\right]\nonumber\\
&&\left.+\left(-\frac{1}{3}\bar{d}'\gamma^\mu_{}d'+\frac{2}{3}\bar{u}'\gamma^\mu_{}u'
-\bar{e}'\gamma^\mu_{}e'\right)\right\}\,.
\end{eqnarray}

As long as the dark photon $A'$ is heavy enough, it can efficiently
decay into the ordinary fermions. The Higgs boson from the dark
Higgs scalar $\eta'$ can decay into the dark photon. The dark scalar
$\eta$ can have a three-body decay mode into the ordinary leptoquark
and diquark scalars (two $\delta$ and one $\omega$) or a two-body
decay mode into the ordinary and dark leptoquark scalars (one
$\delta$ and one $\delta'$).

\section{Predictive dark matter mass and its implication}

The Yukawa couplings (\ref{yukawa}) can tell us the masses of the
dark quarks and charged leptons from the ordinary ones,
\begin{eqnarray}
\label{dqlmass}
&&\frac{m_{d'}^{}}{m_d^{}}=\frac{m_{s'}^{}}{m_s^{}}=\frac{m_{b'}^{}}{m_b^{}}
=\frac{m_{u'}^{}}{m_u^{}}=\frac{m_{c'}^{}}{m_c^{}}=\frac{m_{t'}^{}}{m_t^{}}
=\frac{m_{e'}^{}}{m_{e}^{}}\nonumber\\
&&=\frac{m_{\mu'}^{}}{m_{\mu}^{}}=\frac{m_{\tau'}^{}}{m_{\tau}^{}}
=\frac{\langle\phi'^{}\rangle}{\langle\phi\rangle}\,.
\end{eqnarray}

As we will show later, the dark neutrinos and quarks are unstable
and don't contribute to the relic density of the present universe.
In other words, the dark electron is the unique stable particle in
the dark sector. The dark electron can annihilate into the dark
photon,
\begin{eqnarray}
\sigma_{e'}^{}=\langle\sigma_{e'^{+}_{}e'^{-}_{}\rightarrow
A'A'}^{}v_{\textrm{rel}}^{}\rangle=\frac{\pi \alpha^2_{}}{m_{e'}^2}
~~\textrm{for}~~m_{e'}^{}\gg m_{A'}^{}\,.
\end{eqnarray}
The relic density of the dark electron then can be calculated by
\cite{kt1990}
\begin{eqnarray}
\label{relic} \Omega_{e'}^{}h^2_{}=\frac{1.07\times
10^{9}_{}m_{e'}^{}}{\sqrt{g_\ast^{}}M_{\textrm{Pl}}^{}\sigma_{e'}^{}T_f^{}(\textrm{GeV})}\,,
\end{eqnarray}
where $M_{\textrm{Pl}}^{}\simeq 1.22\times 10^{19}_{}\,\textrm{GeV}$
is the Planck mass, $T_f^{}$ is the freeze-out temperature
\cite{kt1990},
\begin{eqnarray}
\frac{m_{e'}^{}}{T_f}&=& \ln(0.152\,
M_{\textrm{Pl}}^{}m_{e'}^{}\sigma_{e'}^{}/\sqrt{g_\ast^{}})\nonumber\\
&&-\frac{1}{2}\ln[\ln(0.152\,
M_{\textrm{Pl}}^{}m_{e'}^{}\sigma_{e'}^{}/\sqrt{g_\ast^{}})]\,,~~~~~
\end{eqnarray}
while $g_{\ast}^{}=g_{\ast}^{}(T)$ is the number of relativistic
degrees of freedom. Remarkably, the relic density (\ref{relic}) only
depends on one unknown parameter $m_{e'}^{}$, the dark matter mass,
similar to that in some minimal dark matter scenarios where the
exotic dark matter particles only have the SM gauge interactions
\cite{cfs2005}. By inputting $g_{\ast}^{}=90$ \cite{kt1990}, we find
if the dark electron is expected to account for the dark matter
relic density \cite{ade2013}, its mass should have the following
value,
\begin{eqnarray}
\label{dmmass} m_{e'}^{}\simeq 302.1\pm
3.5\,\textrm{GeV}~\textrm{for}~\Omega_{e'}^{}h^2_{}=
0.1199\pm0.0027\,.
\end{eqnarray}

Through the exchange of the dark photon, the dark electron can
scatter off the ordinary nucleon,
\begin{eqnarray}
\sigma_{e'N\rightarrow e'N}&\simeq& \epsilon^2_{}\frac{\pi
\alpha^2_{}\mu_r^2}{m_{A'}^4}
\left[\frac{3Z+(A-Z)}{A}\right]^2_{}\,,\nonumber\\
&\simeq &
10^{-45}_{}\,\textrm{cm}^2_{}\left(\frac{\epsilon}{1.25\times
10^{-7}_{}}\right)^2_{}\left(\frac{\mu_r^{}}{1\,\textrm{GeV}}\right)^2_{}\nonumber\\
&&\times
\left(\frac{1\,\textrm{GeV}}{m_{A'}^{}}\right)^4_{}\left[\frac{3Z+(A-Z)}{A}\right]^2_{}\,,
\end{eqnarray}
which is accessible to the ongoing and future dark matter direct
detection experiments such as the XENON100 \cite{aprile2012} and
XENON1T experiments. Here $Z$ and $A-Z$ are the numbers of proton
and neutron within the target nucleus, while
\begin{eqnarray}
\mu_r&=&\frac{m_{e'}^{} m_N^{}}{m_{e'}^{}+m_N^{}}\simeq
m_{N}^{}\simeq 1\,\textrm{GeV}~~\textrm{for}~~m_{e'}^{}\simeq
302\,\textrm{GeV}\,,\nonumber\\
&&
\end{eqnarray}
is the reduced mass. The dark photon will also mediate a
self-interaction of the dark electron. For example, we can have a
self-interacting cross section,
\begin{eqnarray}
&&\frac{\sigma_{e'^{+}_{}e'^{-}_{}\rightarrow
e'^{+}_{}e'^{-}_{}}^{}}{m_{e'}^{}}
\simeq\frac{4\pi\alpha^2_{}m_{e'}^{}}{m_{A'}^4}\nonumber\\
&\simeq&1.6\times
10^{-42}_{}\,\textrm{cm}^3_{}\left(\frac{m_{e'}^{}}{302\,\textrm{GeV}}\right)
\left(\frac{1\,\textrm{GeV}}{m_{A'}^{}}\right)^4_{}\,,
\end{eqnarray}
which is easy to satisfy the observation limits \cite{rmcgb2007}.

As the dark electron mass is determined, we can fix the dark VEV
$\langle\phi'\rangle$ from Eq. (\ref{dqlmass}),
\begin{eqnarray}
\langle\phi'\rangle&=&\frac{m_{e'}^{}}{m_e^{}}\langle\phi\rangle
\simeq
10^{8}_{}\,\textrm{GeV}\nonumber\\
&&\times \left(\frac{m_{e'}^{}}{302\,\textrm{GeV}}\right)
\left(\frac{0.511\,\textrm{MeV}}{m_e^{}}\right)\left(\frac{\langle\phi\rangle}{174\,\textrm{GeV}}\right)\,.\quad
\end{eqnarray}
Accordingly, the masses of the other dark quarks and charged leptons
should be
\begin{eqnarray}
\begin{array}{lcrclcr}
m_{d'}^{}&=&2.8\,\textrm{TeV}&\textrm{for}&m_{d}^{}&=&4.8\,\textrm{MeV}\,,\\
m_{u'}^{}&=&1.3\,\textrm{TeV}&\textrm{for}&m_{u}^{}&=&2.3\,\textrm{MeV}\,,\\
m_{s'}^{}&=&55\,\textrm{TeV}&\textrm{for}&m_s^{}&=&95\,\textrm{MeV}\,,\\
m_{c'}^{}&=&738.6\,\textrm{TeV}&\textrm{for}&m_{c}^{}&=&1.275\,\textrm{GeV}\,,\\
m_{b'}^{}&=&2.42\times 10^3_{}\,\textrm{TeV}&\textrm{for}&m_{b}^{}&=&4.18\,\textrm{GeV}\,,\\
m_{t'}^{}&=&9.971\times 10^{5}\,\textrm{TeV}&\textrm{for}&m_{t}^{}&=&173.5\,\textrm{GeV}\,,\\
m_{\mu'}^{}&=&61.23\,\textrm{TeV}&\textrm{for}&m_{\mu}^{}&=&105.7\,\textrm{MeV}\,,\\
m_{\tau'}^{}&=&1.029\times
10^{3}_{}\,\textrm{TeV}&\textrm{for}&m_{\tau}^{}&=&1.777\,\textrm{GeV}\,.
\end{array}
\end{eqnarray}

\section{Consequence and fate of dark quarks}

Not only the ordinary quarks but also the dark quarks are related to
the non-perturbative QCD Lagrangian,
\begin{eqnarray}
\mathcal{L}\supset\bar{\theta}\frac{g^2_3}{32\pi^2_{}}G\tilde{G}~~\textrm{with}~~
\bar{\theta}=\theta+\textrm{Arg}\textrm{Det} (M_u^{} M_d^{})\,,
\end{eqnarray}
where $\theta$ is the original QCD phase, while $M_u^{}$ and
$M_d^{}$ are the mass matrices of the up-type and down-type quarks,
\begin{eqnarray}
\mathcal{L}&\supset& -[\bar{d}^{}_L~
\bar{d}'^{}_L]M_d^{}\left[\begin{array}{c}d^{}_R\\
[1mm] d'^{}_R\end{array}\right]-[\bar{u}^{}_L~
\bar{u}'^{}_L]M_u^{}\left[\begin{array}{c}u^{}_R\\
[1mm]
u'^{}_R\end{array}\right]+\textrm{H.c.}\nonumber\\
[1mm] &&\textrm{with}~~ M_{d(u)}^{}=
\left[\begin{array}{cc}y_{d(u)}^{}\langle\phi\rangle&0\\
[1mm] 0&y_{d(u)}^{\dagger}\langle\phi'\rangle\end{array}\right]\,.
\end{eqnarray}
Obviously, the $\theta$-term in the QCD Lagrangian should be zero
because of the parity invariance while the
$[\textrm{Arg}\textrm{Det} (M_u^{} M_d^{})]$-term should be also
trivial due to the real determinants $\textrm{Det}(M_d^{})$ and
$\textrm{Det}(M_u^{})$. We hence can obtain a vanishing strong CP
phase $\bar{\theta}=0$ \cite{bcs1991,lavoura1997,gu2012}.

From Eqs. (\ref{potential}) and (\ref{yukawa}), it is easy to see
that after the dark electromagnetic symmetry breaking, a dark up
quark can decay into a dark electron and four ordinary fermions
through the mediations of the real and/or virtual colored scalars.
For example, we can naively estimate
\begin{eqnarray}
\Gamma_{u'\rightarrow
e'^{+}_{}e^{+}_{}\bar{u}\bar{u}\bar{d}}&\sim&\!
\frac{1}{2^{16}_{}\pi^7_{}}\frac{\langle\eta'\rangle^{2}_{}m_{u'}^{11}}{M_{\delta'}^{4}M_{\delta}^{4}M_{\omega}^{4}}
|\lambda|^2_{}[(y^\dagger_{\delta}y^{}_{\delta})_{11}^{}
\nonumber\\
&&+(y''^\dagger_{\delta}y''^{}_{\delta})_{11}^{}]
[\textrm{Tr}(y^\dagger_{\delta}y^{}_{\delta})+\textrm{Tr}(y''^\dagger_{\delta}y''^{}_{\delta})]\nonumber\\
&&\times
[\textrm{Tr}(y^\dagger_{\omega}y^{}_{\omega})+2\textrm{Tr}(y'^\dagger_{\omega}y'^{}_{\omega})]\nonumber\\
&\simeq&\!\frac{1}{3.5\times
10^{-11}_{}\,\textrm{sec}}\!\left(\frac{\langle\eta\rangle}{7\,\textrm{GeV}}\right)^2_{}
\!\left(\frac{m_{u'}^{}}{1.3\,\textrm{TeV}}\right)^{11}_{}\nonumber\\
&&\times\left(\frac{10\,m_{u'}^{}}{M_{\delta'}^{}}\right)^{4}_{}
\left(\frac{m_{u'}^{}}{M_{\delta}^{}}\right)^{4}_{}
\left(\frac{m_{u'}^{}}{M_{\omega}^{}}\right)^{4}_{}\nonumber\\
&&
\times|\lambda|^2_{}[(y^\dagger_{\delta}y^{}_{\delta})_{11}^{}+(y''^\dagger_{\delta}y''^{}_{\delta})_{11}^{}]
[\textrm{Tr}(y^\dagger_{\delta}y^{}_{\delta})\nonumber\\
&&+\textrm{Tr}(y''^\dagger_{\delta}y''^{}_{\delta})]
[\textrm{Tr}(y^\dagger_{\omega}y^{}_{\omega})+2\textrm{Tr}(y'^\dagger_{\omega}y'^{}_{\omega})]\,.\nonumber\\
&&
\end{eqnarray}
So, the lightest dark quark can have a very short lifetime for a
proper parameter choice.

\section{Neutrino masses and baryon asymmetry}

From the Yukawa couplings (\ref{yukawa}), we can read the mass terms
involving the ordinary neutrinos $\nu_L^{}$, the dark neutrinos
$\nu'^{}_R$ and the singlet fermions $\chi_R^{}$,
\begin{eqnarray}
\mathcal{L}\!\!&\supset&\!\!-\frac{1}{2} [\begin{array}{ccc}
\bar{\nu}^{}_L&\bar{\nu}'^{c}_R&\bar{\chi}^c_R\end{array}]\!\!\left[\begin{array}{ccc}
0&f\langle\Sigma\rangle&h\langle\phi^{}_u\rangle\\
[1mm]
f^T_{}\langle\Sigma\rangle&0&h^\ast_{}\langle\phi'^{}_{}\rangle\\
[1mm] h^T_{}\langle\phi^{}_{}\rangle
&h^\dagger_{}\langle\phi'^{}_{}\rangle &\mu
\end{array}\right]\!\!\left[\begin{array}{c}\nu^{c}_L\\
[1mm]
\nu'^{}_R\\
[1mm] \chi^{}_R\end{array}\right]\nonumber\\
&&+\textrm{H.c.}\,.
\end{eqnarray}
Here the Yukawa coupling matrix $f$ are hermitian due to the parity
symmetry. For
$h\langle\phi'^{}_{}\rangle\gg\mu,h\langle\phi^{}_{}\rangle,f\langle\Sigma\rangle$,
the seesaw mechanism can be applied to give the masses of the
ordinary neutrinos,
\begin{eqnarray}
\label{ilseesaw}
m_\nu^{}&=&f\langle\Sigma\rangle\frac{1}{h^\dagger_{}\langle\phi'^{}_{}\rangle}\mu
\frac{1}{h^\ast_{}\langle\phi'^{}_{}\rangle}f^T_{}\langle\Sigma\rangle
\nonumber\\
&&-
(f\langle\Sigma\rangle\frac{1}{h^\dagger_{}}h^T_{}+h\frac{1}{h^\ast_{}}f^T_{}\langle\Sigma\rangle)
\frac{\langle\phi^{}_{}\rangle}{\langle\phi'^{}_{}\rangle}\nonumber\\
&=&\tilde{f}\frac{1}{\hat{h}}\tilde{\mu} \frac{1}{\hat{h}}
\tilde{f}^T_{}\frac{\langle\Sigma\rangle^2_{}}{\langle\phi'\rangle^2_{}}
- (\tilde{f}+\tilde{f}^T_{})\frac{
\langle\phi^{}_{}\rangle\langle\Sigma^{}_{}\rangle}{\langle\phi'^{}_{}\rangle}\,,
\end{eqnarray}
where we have defined
\begin{eqnarray}
\label{effy} &&\hat{h}=U^{}_{R} h
V^{T}_{R}=\textrm{diag}\{\hat{h}_1^{}\,,~\hat{h}_2^{}\,,~\hat{h}_3^{}\}\,,\nonumber\\
&& \tilde{f}=fU^{\dagger}_R\,,~\tilde{\mu}=V^{}_R \mu V^{T}_R\,.
\end{eqnarray}
The first term of Eq. (\ref{ilseesaw}) is the inverse seesaw
\cite{mv1986}, while the second term is the linear seesaw
\cite{barr2003}.

In this inverse and linear seesaw scenario, a dark neutrino
$\nu'^{}_{Ri}$ and a singlet fermion $\chi_{Ri}^{}$ actually form
two quasi-degenerate Majorana fermions \cite{gs2010},
\begin{eqnarray}
N^{+}_{i}&\simeq&
\frac{1}{\sqrt{2}}(\nu'^{}_{Ri}+\chi^{}_{Ri}+\nu'^{c}_{Ri}+\chi^{c}_{Ri})~~\textrm{with}\nonumber\\
&&M_{N^{+}_i}^{}=\hat{h}_i^{}\langle\phi'^{}_{}\rangle+\frac{1}{2}\tilde{\mu}_{ii}^{}\,,\nonumber\\
N^{-}_{i}&\simeq&
\frac{i}{\sqrt{2}}(\nu'^{}_{Ri}-\chi^{}_{Ri}-\nu'^{c}_{Ri}+\chi^{c}_{Ri})~~\textrm{with}\nonumber\\
&&M_{N^{-}_i}^{}=\hat{h}_i^{}\langle\phi'^{}_{}\rangle-\frac{1}{2}\tilde{\mu}_{ii}^{}\,.
\end{eqnarray}
The Yukawa couplings of the quasi-degenerate Majorana fermions to
the ordinary leptons and Higgs scalar should be
\begin{eqnarray}
\label{eyukawa} \mathcal{L}&\supset& -y_{+}^{}\bar{l}^{}_L\phi
N^{+}_{} -y_{-}^{}\bar{l}^{}_L\phi
N^{-}_{}+\textrm{H.c.}~~\textrm{with}\nonumber\\
&&y_{+}^{}=\frac{1}{\sqrt{2}}(U^{\dagger}_{R}
\hat{h}-\frac{\rho\langle\phi'\rangle}{M^2_\Sigma}\tilde{f})=\frac{1}{\sqrt{2}}(U^{\dagger}_{R}
\hat{h}+\bar{f})\,,\nonumber\\
&&y_{-}^{}=\frac{i}{\sqrt{2}}(U^{\dagger}_{R}
\hat{h}+\frac{\rho\langle\phi'\rangle}{M^2_\Sigma}\tilde{f})=
\frac{i}{\sqrt{2}}(U^{\dagger}_{R} \hat{h}-\bar{f})\,.~~
\end{eqnarray}

Such quasi-degenerate Majorana fermions can accommodate a resonant
leptogenesis to generate the baryon asymmetry in the universe.
Following \cite{pilaftsis1997}, we compute the CP asymmetries from
self-energy corrections,
\begin{eqnarray}
\varepsilon_{N^{\pm}_i}^{}&=&\frac{\Gamma(N^{\pm}_i\rightarrow
l_L^{}\bar{\phi}) -\Gamma(N^{\pm}_i\rightarrow
\bar{l}_L^{}\phi)}{\Gamma(N^{\pm}_i\rightarrow l_L^{}\bar{\phi})
+\Gamma(N^{\pm}_i\rightarrow
\bar{l}_L^{}\phi)}\nonumber\\
&\simeq& \frac{\textrm{Im}\{[(y_{+}^\dagger
y_{-}^{})_{ii}]^2_{}\}}{8\pi(y_{\pm}^\dagger
y_{\pm}^{})_{ii}}\frac{r_{N_i^{}}^{}}{r_{N_i^{}}^2
+\frac{1}{64\pi^2_{}}[(y_{\mp}^\dagger y_{\mp}^{})_{ii}^{}]^2_{}}\,,
\end{eqnarray}
which can be specified by expanding
\begin{eqnarray}
r_{N_i^{}}^{}&=&\frac{M_{N_i^{+}}^{2}-M_{N_i^{-}}^{2}}{M_{N_i^{+}}^{}M_{N_i^{-}}^{}}
\simeq
\frac{2\tilde{\mu}_{ii}^{}}{\hat{h}_i^{}\langle\phi'\rangle}\,,\nonumber\\
(y_{\pm}^\dagger y_{\pm}^{})_{ii}^{}&=& \frac{1}{2}\{\hat{h}_i^2\pm
2 \hat{h}_i^{}\textrm{Re}[(U^{}_R\bar{f})_{ii}^{}]
+(\bar{f}^\dagger_{}\bar{f})_{ii}^{}\}\,,\nonumber\\
\textrm{Im}\{[(y_{\pm}^\dagger y_{\pm}^{})_{ii}^{}]^2_{}\}
&=&\hat{h}^{}_{i}[\hat{h}_i^2 -(\bar{f}^\dagger_{}\bar{f})_{ii}^{}]
\textrm{Im}[(U^{}_R\bar{f})_{ii}^{}]\,.
\end{eqnarray}
When the Majorana fermions $N_i^\pm$ go out of equilibrium, their
CP-violating decays can generate a lepton asymmetry in the ordinary
leptons $l_L^{}$. This lepton asymmetry then will be partially
converted to a baryon asymmetry through the sphaleron processes
\cite{krs1985}. The induced baryon asymmetry can be approximately
described by \cite{kt1990},
\begin{eqnarray}
\eta_B^{}&\simeq& -\frac{28}{79}\times\sum
\frac{\varepsilon_{N_i^{\pm}}^{}\kappa_{N_i^{\pm}}^{}}{g_\ast^{}}
~~\textrm{with}\nonumber\\
&&~~\kappa_{N_i^{\pm}}^{}\simeq\left\{\begin{array}{cl}1 &
\textrm{for}~~K_{N_i^{\pm}}^{}\ll 1\,,\\
[3mm] \frac{0.3}{K_{N_i^{\pm}}^{}(\ln
K_{N_i^{}}^{})^{0.6}_{}}&\textrm{for}~~K_{N_i^{\pm}}^{}\gtrsim
1\,,\end{array}\right.
\end{eqnarray}
where the parameters $K_{N_i^{\pm}}^{}$ are defined as
\begin{eqnarray}
K_{N_i^{\pm}}^{}=\frac{\Gamma_{N_i^\pm}^{}}{2H(T)}\left|_{T=M_{N_i^{\pm}}^{}}^{}\right.\,,
\end{eqnarray}
with $\Gamma_{N_i^\pm}^{}$ being the decay width,
\begin{eqnarray}
\Gamma_{N_i^\pm}^{}&=&\Gamma(N^{\pm}_i\rightarrow l_L^{}\bar{\phi})
+\Gamma(N^{\pm}_i\rightarrow \bar{l}_L^{}\phi)\nonumber\\
&=&\frac{1}{8\pi}(y_{\pm}^\dagger
y_{\pm}^{})_{ii}^{}M_{N_i^\pm}^{}\,,
\end{eqnarray}
and $H(T)$ being the Hubble constant,
\begin{eqnarray}
H(T)=\left(\frac{8\pi^{3}_{}g_{\ast}^{}}{90}\right)^{\frac{1}{2}}_{}
\frac{T^{2}_{}}{M_{\textrm{Pl}}^{}}\,.
\end{eqnarray}

We now show a proper parameter choice can simultaneously result in
the required neutrino masses and baryon asymmetry. For this purpose,
we set
\begin{eqnarray}
\label{parameter1} M_{\Sigma^{}}^{}=
\langle\phi'\rangle=10^{8}_{}\,\textrm{GeV}\,,~~\rho=\langle\phi\rangle\,,
\end{eqnarray}
to give the seesaw-suppressed VEV in Eq. (\ref{vev2}) and the
effective Yukawa couplings in Eq. (\ref{eyukawa}),
\begin{eqnarray}
\langle\Sigma\rangle\simeq -0.3\,\textrm{MeV}\,,~~\bar{f}\simeq
-1.74\times 10^{-6}\,\tilde{f}\,.
\end{eqnarray}
By further inputting,
\begin{eqnarray}
\label{parameter2} \hat{h}_1^{}=10^{-5}_{}\ll
\hat{h}_{2,3}^{}\,,~~\tilde{\mu}_{ij}^{}=10\,\textrm{eV}\,,~~\tilde{f}=\mathcal{O}(0.1-1)\,,
\end{eqnarray}
the neutrino masses (\ref{ilseesaw}) can be dominated by the linear
seesaw,
\begin{eqnarray}
m_\nu^{}\simeq -(
\tilde{f}+\tilde{f}^T_{})\frac{\langle\phi\rangle\langle\Sigma\rangle}{\langle\phi'\rangle}
=0.5\,\textrm{eV} (\tilde{f}+\tilde{f}^T_{})\,.
\end{eqnarray}
The above parameter choice can also induce
\begin{eqnarray}
M_{N_{1}^{\pm}}^{}&=&1\,\textrm{TeV}\ll M_{N_{2,3}^{\pm}}^{}\,,~~
K_{N_{1}^{\pm}}^{}\simeq 700\left(\frac{110.75}{g_\ast^{}}\right)^{\frac{1}{2}}_{}\,,\nonumber\\
\varepsilon_{N_{1}^{\pm}}^{}&\simeq& 0.98\times
10^{-3}\left(\frac{\textrm{Im}[(U^{}_R
\tilde{f})_{11}]}{-0.0154}\right)\,,
\end{eqnarray}
to explain the measured baryon asymmetry \cite{ade2013},
\begin{eqnarray}
\eta^{}_B&\simeq&  3.81\times 10^{-9}_{}\times (0.02205\pm
0.00028)\nonumber\\
&\simeq&  (0.8401\pm 0.0107)\times 10^{-10}_{}\,.
\end{eqnarray}

Note the dark leptoquark scalar and gauge bosons can mediate some
scattering and annihilating processes of the decaying Majorana
fermions. As an example, we check the processes
$N_1^{\pm}N_1^{\pm}\rightarrow d'\bar{d}'$ mediated by the dark
leptoquark $\delta'$ and find,
\begin{eqnarray}
\Gamma_S^{}\sim
[(y'^{\dagger}_{\delta}y'^{}_{\delta})_{11}^{}]^2_{}\frac{T^5_{}}{M_{\delta'}^4}\sim
H(T)\Rightarrow \quad\quad\quad\quad\quad\quad\quad\quad\quad&&\nonumber\\
T\sim 3.7\,\textrm{TeV}
\left[\frac{10^{-6}_{}}{(y'^{\dagger}_{\delta}y'^{}_{\delta})_{11}^{}}\right]^{\frac{2}{3}}_{}
\left(\frac{M_{\delta'}^{}}{10\,m_{e'}^{}}\right)^{\frac{4}{3}}_{}
\left(\frac{g_\ast^{}}{159}\right)^{\frac{1}{6}}_{}\,.&&
\end{eqnarray}
Similarly, the other scattering and annihilating processes can also
decouple at a temperature above the leptogenesis epoch
\cite{mz1992}.

\section{Conclusion}

In the presence of an $SU(2)'^{}_R\times U(1)'^{}_Y$ dark parity
extension of the $SU(2)^{}_L\times U(1)^{}_Y$ ordinary electroweak
theory, the strong CP problem can be solved without resorting to an
axion. In this framework, we consider three gauge-singlet fermions
with small Majorana masses and an $[SU(2)^{}_L\times
SU(2)'^{}_R]$-bidoublet Higgs scalar with seesaw-suppressed VEV to
generate the baryon asymmetry by resonant leptogenesis and the
neutrino masses by inverse and linear seesaw. We also introduce some
colored scalars to mediate a fast decay of the lightest dark quark
after the dark electromagnetic symmetry breaking. The dark electron
can keep stable to account for the dark matter relic density if it
has a determined mass around $302\,\textrm{GeV}$. Benefited from the
$U(1)^{}_Y\times U(1)'^{}_Y$ kinetic mixing, the dark matter
particle can be verified by the dark matter direct detection
experiments .

\textbf{Acknowledgement}: I thank Martin Holthausen, Michael
Ramsey-Musolf and He Zhang for discussions. This work is supported
in part by the Sonderforschungsbereich TR 27 of the Deutsche
Forschungsgemeinschaft.

\end{document}